\documentclass[10 pt,final,journal,letterpaper,oneside,onecolumn]{IEEEtran}
\usepackage{multicol}   
\usepackage{multirow}
\usepackage[pdftex]{graphicx}
\usepackage[lined,boxruled]{algorithm2e}
\usepackage{algorithmic}
\usepackage[small]{caption}
\usepackage{float}
\usepackage{xspace}
\usepackage{fancyhdr}
\usepackage{amssymb,amsmath}
\usepackage{subfig}
\usepackage{textcomp} 
\usepackage{epstopdf}
\usepackage{bbding}
\usepackage{color}
\usepackage[bottom]{footmisc}
\begin{document}

\title{Simultaneous Harvest-and-Transmit Ambient Backscatter Communications under Rayleigh Fading}

\author{Furqan Jameel, Tapani Ristaniemi, Imran Khan, and Byung Moo Lee}

\markboth{Accepted in EURASIP Journal on Wireless Communications and Networking}%
{\MakeLowercase{\textit{et al.}}: Wireless Communications}

\maketitle

\begin{abstract}
Ambient backscatter communications is an emerging paradigm and a key enabler for pervasive connectivity of low-powered wireless devices. It is primarily beneficial in the Internet of things (IoT) and the situations where computing and connectivity capabilities expand to sensors and miniature devices that exchange data on a low power budget. The premise of the ambient backscatter communication is to build a network of devices capable of operating in a battery-free manner by means of smart networking, radio frequency (RF) energy harvesting and power management at the granularity of individual bits and instructions. Due to this innovation in communication methods, it is essential to investigate the performance of these devices under practical constraints. To do so, this article formulates a model for wireless-powered ambient backscatter devices and derives a closed-form expression of outage probability under Rayleigh fading. Based on this expression, the article provides the power-splitting factor that balances the tradeoff between energy harvesting and achievable data rate. Our results also shed light on the complex interplay of a power-splitting factor, amount of harvested energy, and the achievable data rates.
\end{abstract}

\begin{IEEEkeywords}
Ambient backscatter communications, Energy harvesting, Internet of Things (IoT), Smart networking, Wireless-powered communication
\end{IEEEkeywords}

\IEEEpeerreviewmaketitle

\section{Introduction}

The grand vision of the Internet of things (IoT) is quickly turning into reality by bringing everything to the Internet \cite{siow2018analytics,kamalinejad2015wireless}. Latest devices ranging from smartphones to implantable sensors and wearables are claiming to be ``IoT capable''. Although significant improvements have been seen from the design perspective of wireless devices, the objective of connecting everything to the Internet is still a far cry \cite{8581856}. It is because several important challenges arise when ensuring ubiquitous connectivity of devices. {As indicated in \cite{Munir2019}, one of the first challenge is the limited life-cycle of miniature wireless devices. The energy constrained nature of devices becomes an obstacle as the massive amount of data is transferred across an IoT network and the devices are required to be operated in an untethered manner.} Then, there is a requirement of communication reliability which is even more difficult to maintain in large-scale wireless systems \cite{8340813}. The increased reliability most often comes at a cost of increased energy consumption which cannot be regulated by small energy reservoirs of miniature IoT devices. Above all, these devices would need to demonstrate services like ultra-reliable low-latency communications (URLLC), enhanced mobile broadband (eMBB), and massive machine type communications (mMTC)for beyond 5G networks. Resultantly, it has become evident that an ultra low-powered communication paradigm is essential for enabling short-range communication among devices, without compromising the reliability of communications \cite{kamalinejad2015wireless,liu2018optimal}.

Of late, backscatter communication has gathered the attention of the researchers as a key enabling technology for connecting IoT devices. Backscatter communication allows radio device to transmit their data by reflecting and modulating an incident radio frequency (RF) signal. It adapts the antenna impedance mismatch in order to change the reflection coefficient. Using the received RF energy, backscatter devices harvest a fraction of energy for circuit operations \cite{boyer2014invited}. It is worth highlighting that the backscatter devices do not require oscillators for generating carrier signals as they get the carrier waves from the dedicated RF source. In fact, the ultra-low power nature of a backscatter transmitter (i.e., below 1 mW \cite{kellogg2014wi}), shows promise for a very long life-cycle (i.e., 10 years) with an on-chip battery. Since the harvested energy from an RF source typically ranges from 1mW to 10s of mW, the low power consumption of backscatter devices is a perfect match for RF energy harvesting \cite{lu2015wireless}. 

Besides the obvious advantages of conventional backscatter communications, there are few limitations of these devices. The backscatter devices require a dedicated RF source for transmission of carrier waves. Even though this model has been adopted in radio frequency identification (RFID) tags used in libraries and grocery stores, the power budget of these communication models may not be suitable for energy constrained IoT devices \cite{liu2018optimal,8417660}. {Additionally, the centralized nature of these communication models is also a hurdle in paving the way for large-scale deployment of IoT networks. The distributed architecture of IoT networks favors the deployment of decentralized RF sources that can be accessed anytime. Besides this, energy harvesting through wireless power transmission can extend the life cycle of the IoT networks with little changes in hardware implementations \cite{zhao2017exploiting,chang2019distributed}} 

To overcome the above-mentioned limitations, a new backscatter paradigm has emerged that is called ambient backscatter communication \cite{liu2013ambient}. An ambient backscatter transmitter uses ambient RF signals in order to perform in a battery-free manner. More specifically, the ambient RF signals are used for backscattering and energy harvesting. This flexibility allows the cost-effective deployment of ambient backscatter devices while avoiding dependence on a particular RF source \cite{han2017wirelessly}. However, owing to the novelty of the technology, the study of ambient backscatter communications is still at its nascent stage. A variety of network challenges and data communication issues arise that require further exploration. Furthermore, limited theoretical knowledge of ambient backscatter communication demands new dimensions for performance evaluation of the network.

{Motivated by the aforementioned observations, we perform the analysis of backscatter communication under Rayleigh fading. Specifically, our contribution is two-fold:
\begin{itemize}
\item Derivation of closed-form expression of outage probability for wireless-powered devices operating under Rayleigh fading.
\item Derivation of the power-splitting factor that balances the tradeoff between energy harvesting and achievable data rate. 
\end{itemize}}

The remainder of the paper is organized as follows. Section 2 discusses the related work on conventional backscatter and ambient backscatter communications. In Section 3, a detailed description of the system model is provided. Section 4 provides the performance analysis while Section 5 discusses the numerical results. Finally, Section 6 provides key findings and conclusions.

\section{Related Work}

Backscatter communication has been considered from different aspects in wireless networks \cite{van2018ambient}. The authors of \cite{lu2018wireless} employed backscatter communication to enable device-to-device communications. Besides this, several detection schemes for backscatter communication systems are proposed in \cite{yang2017cooperative,wang2016ambient,qian2017noncoherent}. A detector that does not require the channel state information (CSI) was constructed using a differential encoder in \cite{wang2016ambient}. Specifically, they developed a model and derived optimal detection and minimum bit-error-rate (BER) thresholds. Moreover, the expressions for lower and upper bounds on BER were also derived that were corroborated through simulation results. A joint-energy detection scheme is proposed in \cite{qian2017noncoherent} that requires only channel variances rather than specific CSI. The same authors provided a study of BER computation, optimal and suboptimal detection, and blind parameter acquisition. The non-coherent signal detection outperformed the conventional techniques in terms of detection accuracy and computation complexity. A successive interference cancellation (SIC) based detector and a maximum-likelihood (ML) detector with known CSI are presented in \cite{yang2017cooperative}, to recover signals not only from readers but also from RF sources. In addition to this, the authors derived BER expressions for the ML detector. It was shown that the backscatter signal can significantly enhance the performance of the ML detector as compared to conventional single-input-multiple-output (SIMO) systems.

Capacity and outage performance analysis for ambient backscatter communication systems was studied in \cite{darsena2017modeling,kang2017riding,zhang2017outage,zhao2018outage}. The authors of \cite{darsena2017modeling} analyzed the channel capacity over orthogonal frequency division multiplexing (OFDM) signals. The ergodic capacity optimization problem at the reader with SIC was investigated by the authors of \cite{kang2017riding}. Specifically, the authors jointly considered the transmit source power and the reflection coefficient and improved the ergodic capacity. For ambient backscatter communication systems, the BER of an energy detector was derived and the BER-based outage probability was obtained in \cite{zhang2017outage}. {Zhao \emph{et al}. in \cite{zhao2018outage}, the effective distribution of signal-to-noise ratio (SNR) was derived and the SNR-based outage probability was evaluated over real Gaussian channels.} 

More recently, the authors in \cite{hoang2017overlay} investigated a cognitive radio network having ambient backscatter communication. In particular, it was considered that a wireless-powered secondary user can either harvest energy or adopt ambient backscattering from a primary user on transmission. A time allocation problem was developed in order to maximize the throughput of the secondary user and to obtain the optimal time ratio between energy harvesting and ambient backscattering. Reference \cite{kim2017hybrid} introduced a hybrid backscatter communication scheme as an alternative access scheme for a wireless-powered transmitter. Specifically, when the ambient RF signals were not sufficient to support wireless-powered communications, the transmitter can choose between bistatic backscattering or ambient backscattering based on a dedicated carrier emitter. A throughput maximization problem was formulated to find the optimal time allocation for the hybrid backscatter communication operation. Both \cite{hoang2017overlay} and \cite{kim2017hybrid} studied a deterministic scenarios. 

\section{Experimental System Model Design}

\begin{figure*}[htp]
\centering
\includegraphics[trim={0 0cm 0 0cm},clip,scale=.5]{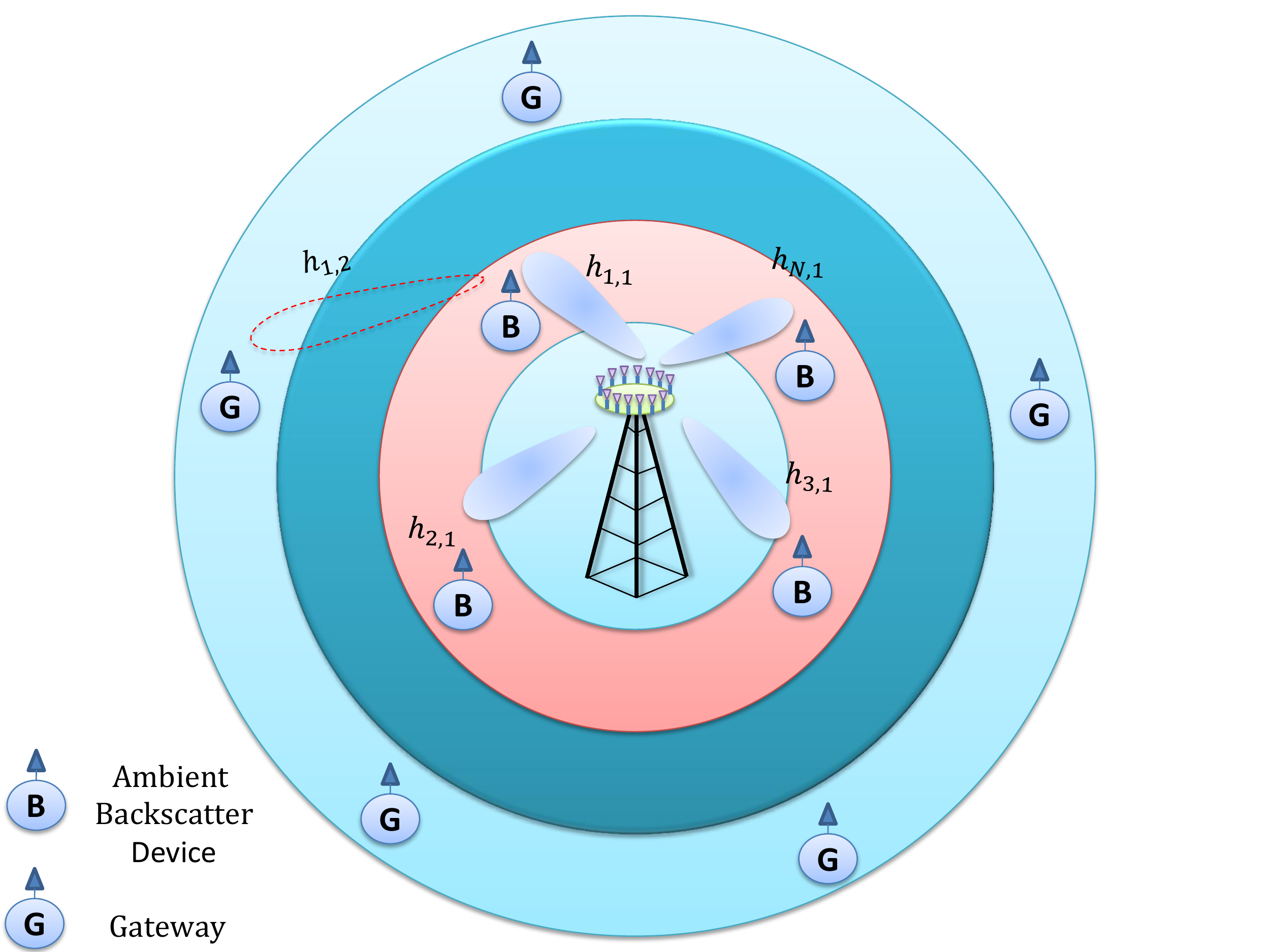}
\caption{System model.}
\label{fig.1}
\end{figure*}

Let us consider an IoT network that consists of $N$ number of ambient backscatter devices. These ambient backscatter devices are considered to be powered by ambient RF source. {This consideration is under the assumption that ambient RF sources (like Radio signals, TV signals and WiFi signals) are abundant in the environment.} These backscatter devices use the harvested energy from the ambient RF signals and transmit their data to the gateway as shown in Figure \ref{fig.1}.

\begin{figure*}[htp]
\centering
\includegraphics[trim={0 0cm 0 0cm},clip,scale=.5]{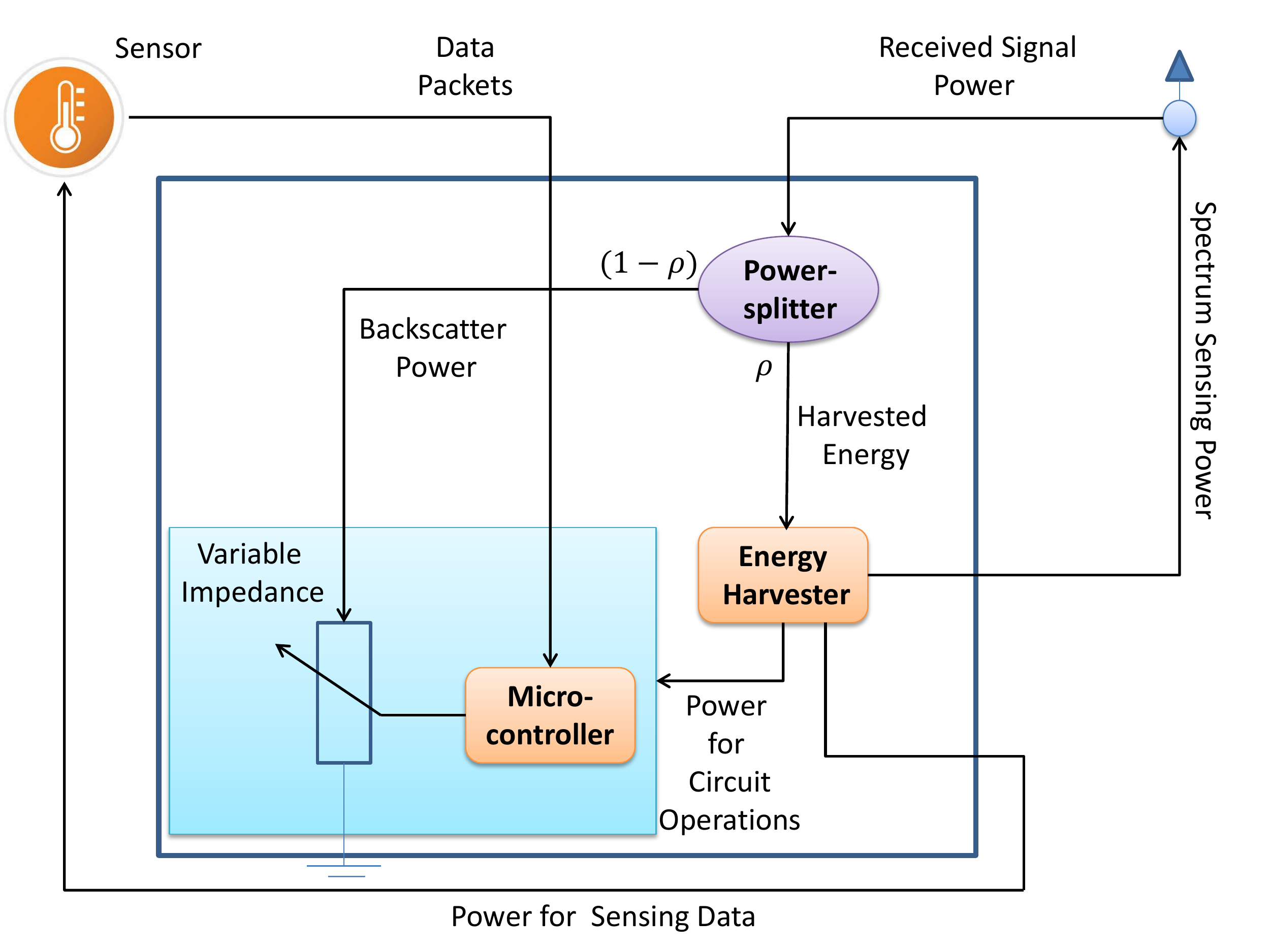}
\caption{Circuit design of the ambient backscatter device.}
\label{fig.2}
\end{figure*}

A typical ambient backscatter device has three major operations, i.e., spectrum sensing, energy harvesting, and data exchange. Inspired by \cite{liu2018optimal}, we consider the circuit model of an ambient backscatter device is shown in Figure \ref{fig.2}. Here, the main purpose of the spectrum sensor is to detect suitable ambient RF signals, whereas, the energy harvesting circuit enables the backscatter devices to operate a self-sustainable manner. This self-sustainability is essential for IoT networks as they are expected to operate with minimum human intervention. When the device is in operation mode, the spectrum sensing is performed in order to detect RF signal with large power. Afterward, the detected signal is employed for either backscatter communication or energy harvesting. The analog-to-digital converter (ADC) uses the harvested energy and convert it into direct current that is utilized by other modules including a microcontroller. The microcontroller performs multiple communication operation including processing the information and matching the impedance of antenna for better reception of RF signals. {We consider that the amount of energy consumed by energy harvester is negligible \cite{liu2018optimal} and satisfies the following condition}

\begin{align}
E_{h}\ge E_{b}+E_{s}+E_{m}.
\end{align}

In the above expression $E_{h}, E_{b},E_{s},E_{m}$ denotes the harvested energy, energy consumed for backscatter communication, energy consumed for spectrum sensing and the energy consumed by micro-controller/ sensor for data gather and processing. {Some of the key symbols used throughout this paper are provided in Table \ref{tabu}.}

\begin{table}[h]
\centering
\begin{tabular}{|l|l|}
\hline
\textbf{Symbol} & \textbf{Definition} \\
\hline
$E_{h}$ & Harvested energy \\
\hline
$E_{b}$ & Energy consumed for backscatter communication \\
\hline
$E_{s}$ & Energy consumed for spectrum sensing \\
\hline
$E_{m}$ & Energy consumed by micro-controller/ sensor \\
\hline
$\alpha $ & Compressive sensing duration \\
\hline
$\rho$ & Power-splitting factor \\
\hline
$\beta$ & Reflection coefficient of the backscatter devices \\
\hline
$\theta$ & Path loss exponent \\
\hline
$N_{0}$ & AWGN variance \\
\hline
$\eta$ & Energy conversion efficiency \\
\hline
$M$ & Number of wideband signals \\
\hline
$e$ & Energy consumed for each sample \\
\hline
$\varphi$ & Threshold of required data rate \\
\hline
$\psi$ & Energy threshold for operation of the backscatter device \\
\hline
$f$ & Sampling rate \\
\hline
$P_{b}$ & Amount of circuit power consumed during backscattering \\
\hline
\end{tabular}
\caption{{Common symbols used in the article.}}
\label{tabu}
\end{table}

\begin{figure*}[htp]
\centering
\includegraphics[trim={0 2cm 0 5cm},clip,scale=.5]{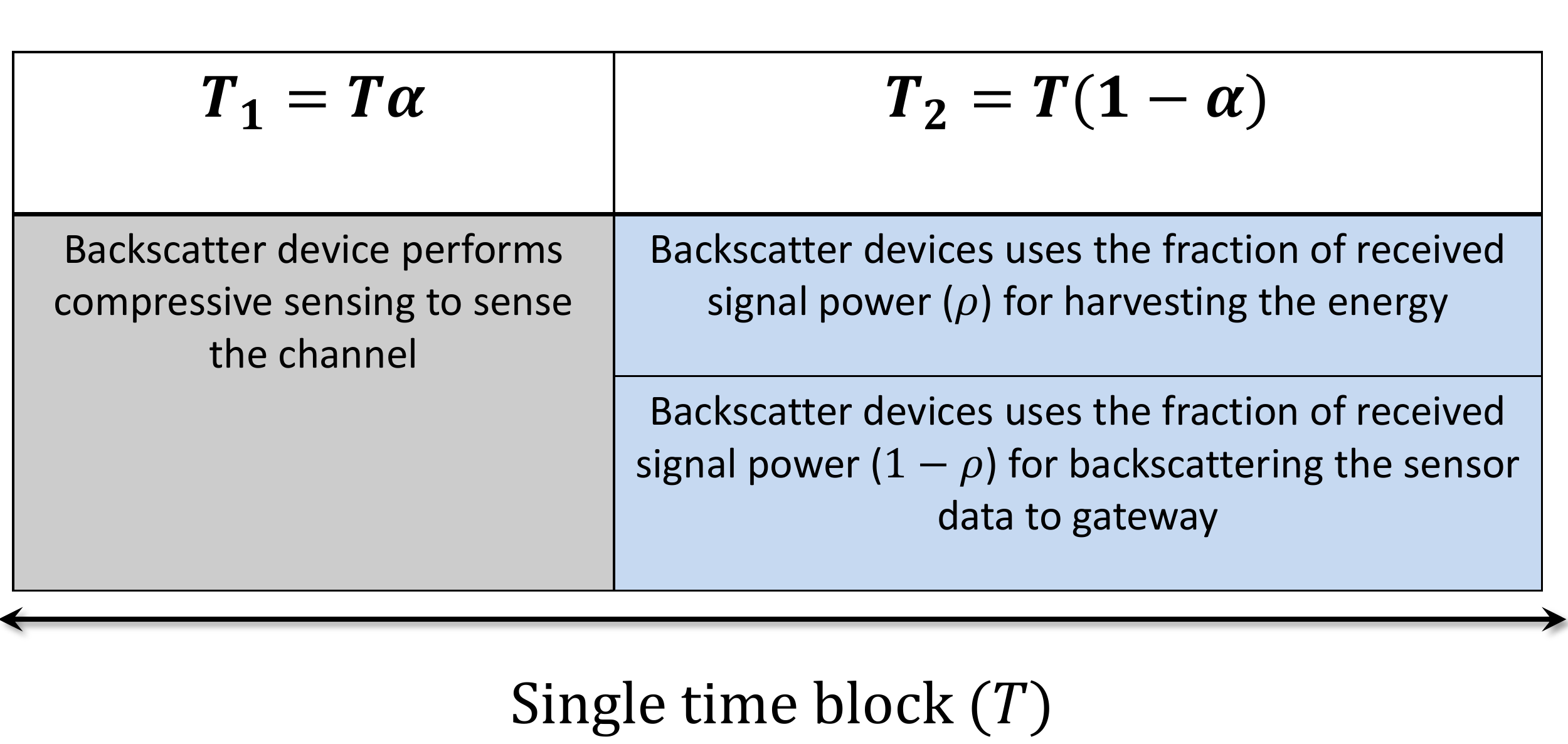}
\caption{Time schedule and power splitting.}
\label{fig.3}
\end{figure*}

We now characterize the energies harvested and consumed during one time
slot. We consider that compressive sensing is performed in each time slot. Thus, the time slot $T$ is divided into phases, i.e., 
compressive sensing duration (denoted as $\alpha $) and energy 
harvesting/ backscattering duration (denoted as ($1-\alpha )$). 
After compressive sensing, the received signal at the device is divided into two streams of power. The first part is used for energy harvesting while the other part is used for performing backscattering operation. 
This separation is performed with a factor $\rho $, where $0<\rho \le 
1$. A graphical representation of an interplay of $\rho $ and $\alpha $ 
is provided in Figure \ref{fig.3}. Assuming that an $i$-th backscatter device 
detects an ambient RF, then the received signal at the device is given 
as
%%%%%%%%%%%%%%%%%%%%%%%%%%%%%%%%%%%%%%

%%%%%%%%%%%%%%%%%%%%%%%%%%%%%%%%%%%%%%

\begin{align}
y_{i,1}=\sqrt{\frac{\beta P}{P_{l,1}}}h_{i,1}s_{1}+n_{i,1},
\end{align}
where $y_{i,1}$ is the received signal, $s_{1}$ denotes the 
normalized signal, $P$ represents the transmit power, and $P_{l,1}=d_1^\theta$ 
is the path loss experienced by the backscatter device and $\theta$ is the path loss exponent. {Furthermore, $
h_{i,1}$ represents the channel gain between the ambient RF source and 
backscatter device which is assumed to be Rayleigh faded, $n_{i,1}$ is the zero mean additive white Gaussian 
noise (AWGN) with $N_{0}$ variance while $\beta $ is the 
reflection coefficient of the backscatter devices.} The harvested energy 
is then denoted as

\begin{align}
E_{h,i}=\frac{\rho \eta (1-\alpha )T\beta \Omega_1 \vert h_{i,1}\vert 
^{2}}{P_{l,1}},
\label{eq6}
\end{align}
where $\Omega _{1}=\frac{P}{N_{0}}$, $\rho $ represents the fraction 
of power used for energy harvesting, $\eta $ is the energy conversion 
efficiency that is considered to be same for all the backscatter devices 
as they employ same circuitry. 

The amount of energy consumed by the compressive sensing module is a linear multiplication of the number of samples and sampling rate. More 
specifically, it can be represented as

\begin{align}
E_{s}=\alpha fMeT,
\end{align}
where $M$ is the number of wideband signals that have been detected 
during the phase of spectrum sensing, $f$ is the sampling rate, and $
e$ is the energy consumed for each sample. 

The amount of energy consumed the backscattering module can be 
represented in terms of circuit power as

\begin{align}
E_{b}=(1-\alpha )P_{b}T,
\end{align}
where $P_{b}$ is the amount of circuit power consumed during 
backscattering phase. For the sake of simplicity and without loss of 
generality, we consider that the power consumed by micro-controller is 
fixed.

As a result of backscattering, the received message at the gateway can 
be written as

\begin{align}
y_{i,2}=\sqrt{\frac{(1-\rho )\beta P_{b}}{P_{l,2}}}h_{i,2}s_{i,2}+n_{i,2},
\end{align}
where $y_{i,2}$ is the received signal at the gateway, $s_{i,2}$ 
denotes the normalized signal sent by the $i$-th backscattering 
device, $P$ represents the transmit power, and $P_{l,2}=d_2^\theta$ is the path 
loss between backscatter device and the gateway. {Furthermore, $h_{i,2}
$ represents the Rayleigh faded channel gain between the backscatter device and the 
gateway, $n_{i,2}$ is the zero mean AWGN with zero mean and $N_{0}$ 
variance.}

\section{Performance Analysis and Methodology}

In this section, we derive the communication outage and power shortage probabilities of the backscatter devices. Based on these probabilities, 
we aim to find the balancing value of the $\rho $.

\subsection{Outage Performance}
Using the Shannon capacity formula, the achievable sum rate at the gateway can be written as

\begin{align}
R_{sum}=\sum_{i=1}^{N}{R_{i}},
\end{align}
where $R_{i}$ is the achievable rate of $i$-th backscattering 
device which is given as

\begin{align}
R_{i}=(1-\alpha )BT\log _{2}\left\{ 1+\frac{(1-\rho )\beta \Omega _{2} 
\vert h_{i,2}\vert ^{2}}{P_{l,2}}\right\}, 
\end{align}
where $\Omega _{2}=\frac{P_{b}}{N_{0}}$.

The outage probability of achievable rate can occur due to the following two 
conditions

\begin{enumerate}
\item If the harvested energy is below the energy required for operations of backscatter device.
\item If the achievable rate is below the required rate at the gateway.
\end{enumerate}

Thus, using the total probability theorem, the outage probability can be 
written as 

\begin{align}
P_{out}&=\Pr(R_{i}<\varphi |E_{h,i}<\psi )\Pr(E_{h,i}<\psi 
) \nonumber \\
&+\Pr(R_{i}<\varphi |E_{h,i}>\psi )\Pr(E_{h,i}>\psi ),
\label{eq1}
\end{align}
where $\varphi $ represents the threshold of required data rate and $\psi =E_{b}+E_{s}+E_{m}$ is the energy threshold for operation of the backscatter device. 

From the above equation, we note that if the harvested energy is below the threshold, then the backscatter device would not be able to transfer any data to the gateway. In this case, the probability that the rate falls below a required threshold would always be 1. Thus, we can write

\begin{align}
\Pr(R_{i}<\varphi |E_{h,i}<\psi )=1.
\label{eq2}
\end{align}

The probability that the harvested energy would fall below a specified threshold can be written as

\begin{align}
\Pr(E_{h,i}<\psi )=\Pr\left(\frac{\rho \eta (1-\alpha )T\beta \Omega 
_{1}\vert h_{i,1}\vert ^{2}}{P_{l,1}}<\psi \right).
\end{align}

After some simplifications, it can be represented as

{\begin{align}
\Pr(E_{h,i}<\psi )&=\Pr\left(\vert h_{i,1}\vert ^{2}<\frac{P_{l,1}\psi }{\rho 
\eta (1-\alpha )T\beta \Omega _{1}}\right) \nonumber \\
&=1-\exp\left\{-\frac{P_{l,1}\psi }{\bar{\gamma_1}\rho \eta (1-\alpha )T\beta \Omega _{1}}\right\}.
\label{eq3}
\end{align}}

In contrast, the probability of energy harvesting increasing beyond the threshold can be represented as

\begin{align}
\Pr(E_{h,i}>\psi )=\exp\left\{-\frac{P_{l,1}\psi }{\bar{\gamma }_{1}\rho \eta 
(1-\alpha )T\beta \Omega _{1}}\right\},
\label{eq4}
\end{align}
{where $\bar{\gamma_1 }$ is the average channel gain between RF source 
and the backscattering device.} Let us now consider the case when the 
harvested energy is greater than $\psi $. In this case, the 
probability that the achievable data rate falls below a pre-determined 
threshold can be written as 

\begin{align}
&\Pr(R_{i}<\varphi |E_{h,i}>\psi )=\Pr\biggl[(1-\alpha )BT \nonumber \\
&\times \log _{2}\left\{
1+\frac{(1-\rho )\beta \Omega _{2} \vert h_{i,2}\vert 
^{2}}{P_{l,2}}\right\} <\varphi |E_{h,i}>\psi\biggr].
\end{align}

After some straightforward simplifications, we obtain

\begin{align}
\Pr(R_{i}<\varphi |E_{h,i}>\psi )&=\Pr\left(\vert h_{i,2}\vert 
^{2}<\frac{P_{l,2}(2^{\frac{\varphi }{(1-\alpha )BT}}-1)}{(1-\rho )\beta 
\Omega _{2}}\right) \nonumber \\
&=1-\exp\left\{-\frac{P_{l,2}(2^{\frac{\varphi }{(1-\alpha 
)BT}}-1)}{\bar{\gamma }_{2}(1-\rho )\beta \Omega _{2}}\right\},
\label{eq5}
\end{align}
where $\bar{\gamma }_{2}$ is the average channel gain between 
backscattering device and the gateway. {Substituting the Eqs (\ref{eq2}), (\ref{eq3}), (\ref{eq4}), and (\ref{eq5}) in (\ref{eq1}), we obtain
\begin{align}
P_{out}&=1-\exp\left\{-\frac{P_{l,1}\psi }{\bar{\gamma }_{1}\rho \eta 
(1-\alpha )T\beta \Omega _{1}}\right\}+ \exp\left\{-\frac{P_{l,1}\psi }{\bar{\gamma }_{1}\rho \eta 
(1-\alpha )T\beta \Omega _{1}}\right\} \nonumber \\
&\times \left[1-\exp\left\{-\frac{P_{l,2}(2^{\frac{\varphi }{(1-\alpha 
)BT}}-1)}{\bar{\gamma }_{2}(1-\rho )\beta \Omega _{2}}\right\}\right].
\label{erqe}
\end{align}}

{After solving \ref{erqe}, we have
\begin{align}
P_{out}=1-\exp\biggl(-\frac{P_{l,2}(2^{\frac{\varphi }{(1-\alpha 
)BT}}-1)}{\bar{\gamma }_{2}(1-\rho )\beta \Omega _{2}}-\frac{P_{l,1}\psi 
}{\bar{\gamma }_{1}\rho \eta (1-\alpha )T\beta \Omega _{1}}\biggr).
\end{align}}

\subsection{Balancing communication outage and power shortage}

In this section, we aim to find the values of $\rho $ that balances the tradeoff between communication outage and power shortage. In particular, 
we note that different values of $\rho $ have a different impact on communication outage and power shortage. From (\ref{eq6}), we can observe that the amount of energy harvested is the increasing function of $\rho $. In other words, as the value of $\rho $ increases, the amount of harvested energy also increases, whereas, it decreases with a 
decrease in the value of $\rho $. In contrast, the achievable rate of 
any $i$-th backscattering device is a decreasing function of $\rho $. Since the achievable rate is dependent on the received SNR, 
therefore, increasing the value of $\rho $ results in increasing the 
SNR while a reduction in $\rho $ causes an increase in the values of 
SNR which in turn increases the achievable rate.

From the above arguments, we can observe that the balancing value of $
\rho $ can be found by solving the energy harvesting and SNR expressions 
simultaneously. Thus, we can write
\begin{align}
\frac{\rho \eta (1-\alpha )T\beta \Omega _{1}\vert h_{i,1}\vert 
^{2}}{P_{l,1}}=\frac{(1-\rho )\beta \Omega _{2} \vert h_{i,2}\vert 
^{2}}{P_{l,2}}.
\end{align}

{After cross multiplication and taking least common multiple, we obtain the $\rho ^{*}$ as
\begin{align}
\rho ^{*}=\frac{\vert h_{i,2}\vert ^{2}\Omega _{2}P_{l,1}}{\vert 
h_{i,2}\vert ^{2}\Omega _{2}P_{l,1}+\eta (1-\alpha )T\beta \Omega 
_{1}\vert h_{i,1}\vert ^{2}P_{l,2}}.
\end{align}}

From the above expression, we can observe that $\rho ^{*}$ is inversely proportional to the $\Omega _{1}$. Moreover, if $P_{l,1}=P_{l,2}$, 
then the balancing value of $\rho ^{*}$ is halved. We also note that 
the value of $\rho ^{*}$ increases with an increase in $\alpha $ 
indicating the direct relationship between $\rho ^{*}$ and $\alpha$.

\section{Results and Discussions}

In this section, we provide results and relevant discussion on the above-mentioned analysis. {Unless mentioned otherwise, following parameters have been used for generating simulation and analytical results: $\eta=0.5$, $B=1MHz$, $\beta=0.5$, $d_1=d_2=$5m, $\varphi=$2kbps, $\theta=2$, and $\rho=0.3$.} 

\begin{figure*}[htp]
\centering
\includegraphics[trim={0 0cm 0 0cm},clip,scale=.5]{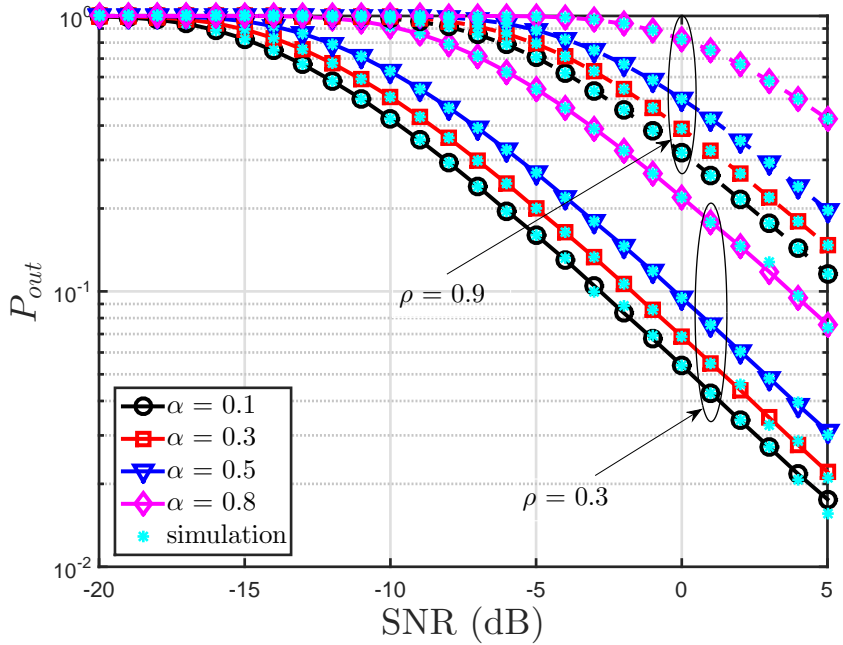}
\caption{Outage probability as a function of SNR.}
\label{fig.5}
\end{figure*}

Figure \ref{fig.5} illustrates the outage probability as a function of increasing values of SNR. It can be seen that the outage probability decreases with an increase in the SNR. However, the impact of $\alpha$ on $P_{out}$ is different for different values of SNR. Specifically, we observe that an increase in the $\alpha$ results in an increase in the outage probability. It is because with an increase in $\alpha$ provides more time for compressive sensing and less time for energy harvesting and backscattering. On the other hand, an increase in $\rho$ causes an increase in outage probability. This result is caused by allocating more fraction of received power for energy harvesting and less for performing backscatter communications. In addition, the simulation results closely follow the analytical curves which indicates the validity of our theoretical model.

%%%%%%%%%%%%%%%%%%%%%%%%%%%%%%%%%%%%%%

%%%%%%%%%%%%%%%%%%%%%%%%%%%%%%%%%%%%%%
\begin{figure*}[!htp]
\centering
\begin{tabular}{c}
\includegraphics[trim={0 0cm 0 0cm},clip,scale=.5]{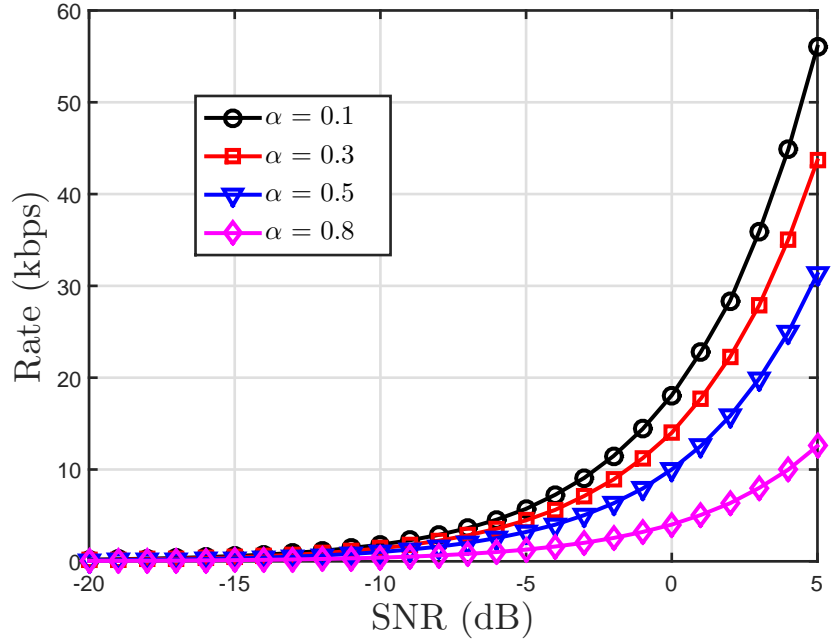} \\
(a) \\
\includegraphics[trim={0 0cm 0 0cm},clip,scale=.5]{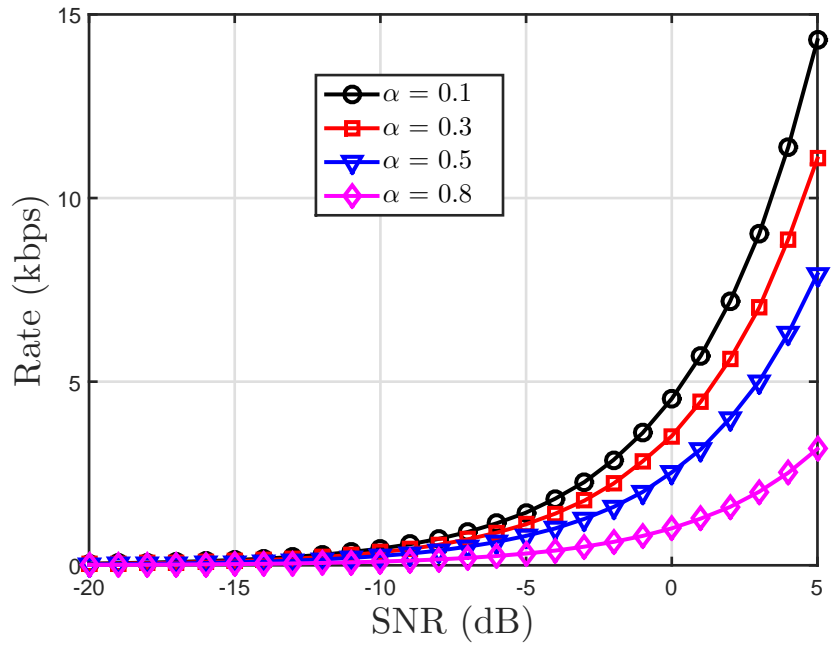} \\
(b) \\
\end{tabular}
\caption{Achievable rate against different values of $\alpha$ where (a) $d_1=d_2=5$m, (b) $d_1=d_2=10$m.}
\label{fig.6}
\end{figure*}

Figure \ref{fig.6} (a) shows the achievable rate as a function of increased SNR. As anticipated by the analytical expression, the increase in SNR improves the achievable rate. However, an increase in $\alpha$ decreases the achievable rate. In fact, the impact of $\alpha$ becomes more prominent at higher values of SNR showing a rapid rise in the curves. Where Figure \ref{fig.6} (a) is plotted for $d_1=d_2=5$m, the curves of Figure \ref{fig.6} (b) are plotted against $d_1=d_2=10$m. This increase in distance has a critical impact on the achievable rate. In particular, for the same values of SNR and $\alpha$ (e.g., SNR=0 dB and $\alpha$=0.1), the achievable rate drops from 20 kbps to 5 kbps when the distance is increased. 

\begin{figure*}[htp]
\centering
\includegraphics[trim={0 0cm 0 0cm},clip,scale=.5]{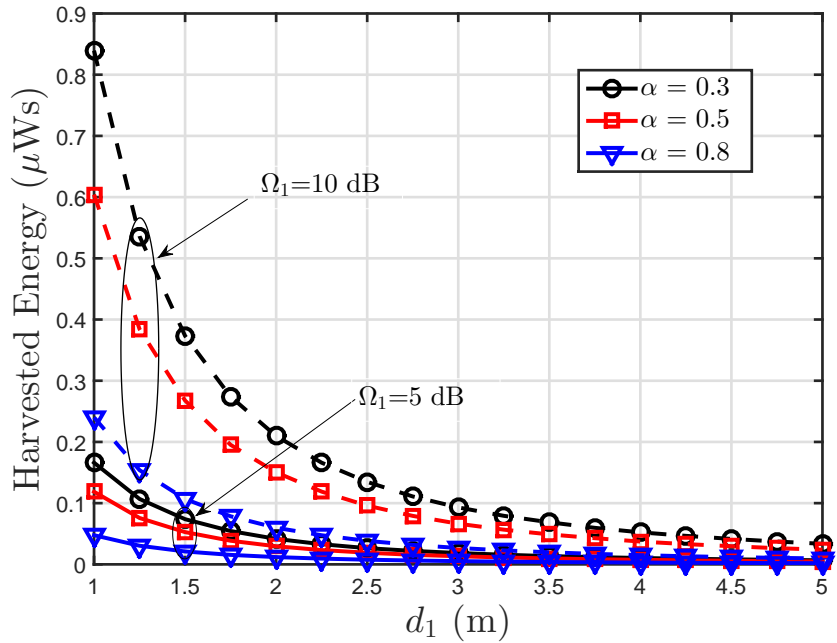}
\caption{Harvested energy against increasing values of $d_1$.}
\label{fig.4}
\end{figure*}

Figure \ref{fig.4} plots the harvested energy against increasing values of $d_1$. Indeed, this results highlight the significance of distance between the RF source and the backscatter device. It can be seen that an increase in $d_1$ results in decreasing the harvested energy. Additionally, the increasing values of $\alpha$ decrease the harvested amount of energy due to compressive sensing. This decrease in harvested energy, against different values of $\alpha$, is less prominent when $\Omega_1=5$ dB. This indicates that the time scheduling is more effective for large transmit power of the ambient RF source. Furthermore, this increase in $\Omega_1$ allows devices to harvest power up to a significantly larger distance which influences the life-cycle of devices. 

%%%%%%%%%%%%%%%%%%%%%%%%%%%%%%%%%%%%%%

%%%%%%%%%%%%%%%%%%%%%%%%%%%%%%%%%%%%%%
\begin{figure*}[htp]
\centering
\begin{tabular}{c}
\includegraphics[trim={0 0cm 0 0cm},clip,scale=.5]{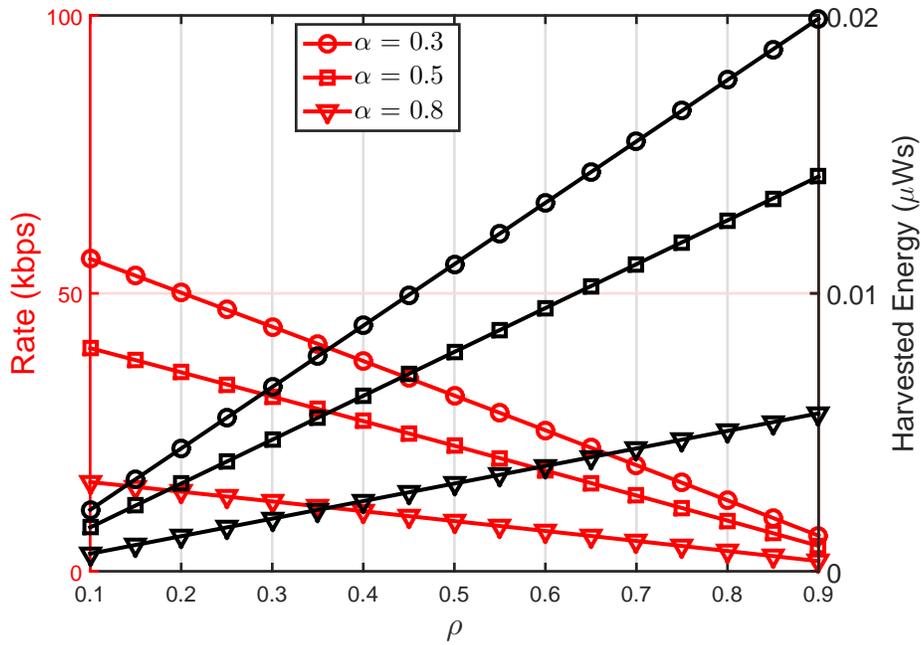} \\
(a) \\
\includegraphics[trim={0 0cm 0 0cm},clip,scale=.5]{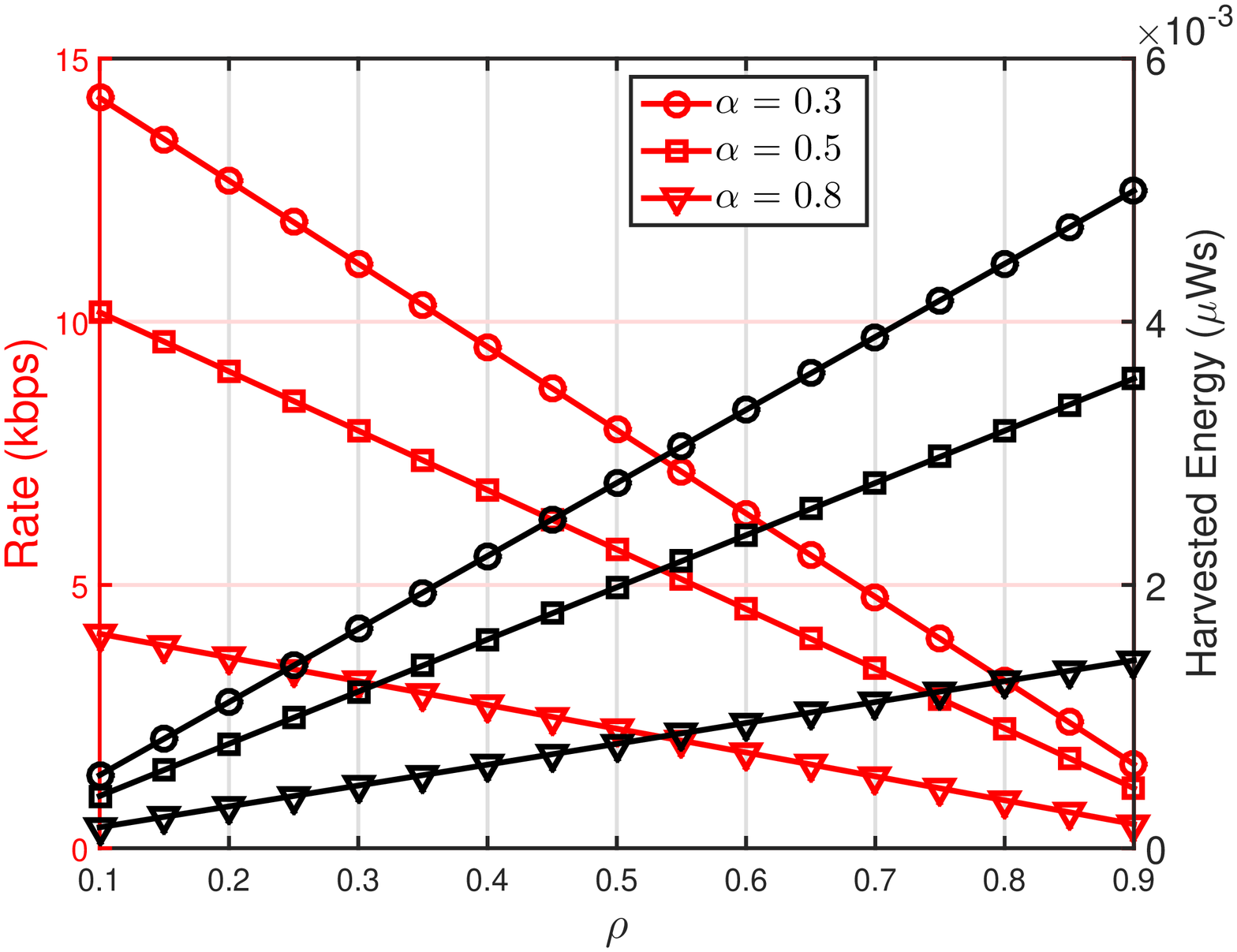} \\
(b) \\
\end{tabular}
\caption{Achievable rate and harvested energy versus increasing values of $\rho$, where $\eta=0.3$ and (a) $d_1=d_2=5$m, (b) $d_1=d_2=10$m.}
\label{fig.7}
\end{figure*}

Figure \ref{fig.7} (a) demonstrates the tradeoff between harvested energy and achievable rate. We have plotted different curves of achievable rate and harvested energy against the increasing values of $\rho$. It can be observed that an increase in $\rho$ causes an increase in the amount of harvested energy while simultaneously reducing the achievable rate. Since the value of $\alpha$ influences both rate and harvested energy, the lower values of $\alpha$ decreases the converging point of the curves of rate and energy curves. Similar trends can be shown in \ref{fig.7} (b), however, the converging point of the curves now shift towards the right-hand side while reducing both the harvested energy and rate. This trend can be attributed to the increase in $d_1$ and $d_2$. This shift in balancing point shows that a higher value of $\rho$ is required with an increase in distance. This also indicates that energy harvesting becomes a critical factor when the distance is increased between ambient RF source and the device and that between device and gateway.

\section{Conclusion}

Ambient backscatter communications provide virtually endless opportunities to connect wireless devices. We anticipate that wearable devices, connected homes, industrial internet, and miniature embeddable are some of the areas where ambient backscatter communications would be adapted to provide pervasive connectivity. Thus, to better analyze the utility of these low-powered devices, this article has provided a comprehensive analysis of ambient backscattering model from the perspective of achievable data rates and the amount of harvested energy. In addition to deriving closed-form expressions of outage probability and balancing power-splitting factor, we have shown that the distance between ambient RF source and the device plays a critical role in determining the life-cycle of devices and the outage probability at the gateway. In fact, we have demonstrated that an increase in distance shifts the balancing power-splitting point to the right-hand side. Besides this, we have observed that when the distance is increased from 5m to 10m against fixed values of SNR and $\alpha$, the achievable rate at gateway drops from 20 kbps to 5 kbps. These results can act as a fundamental building block for designing and large-scale deployment of ambient backscatter devices in the future.
%%%%%%%%%%%%%%%%%%%%%%%%%%%%
\section*{Acknowledgment}

%%%%%%%%%%%%%%%%%%%%%%%%%%%%
\ifCLASSOPTIONcaptionsoff
  \newpage
\fi
%%%%%%%%%%%%%%%%%%%%%%%%
\bibliographystyle{IEEEtran}
\bibliography{Ref}

%%%%%%%%%%%%%%%%%%%%%%%%
\begin{IEEEbiographynophoto}{Furqan Jameel}
received his BS in Electrical Engineering (under ICT R\&D funded Program) in 2013 from the Lahore Campus of COMSATS Institute of Information Technology (CIIT), Pakistan. In 2017, he received his Master's degree in Electrical Engineering (funded by prestigious Higher Education Commission Scholarship) at the Islamabad Campus of CIIT. In 2018, he visited Simula Research Laboratory, Oslo, Norway. Currently, he is a researcher at the University of Jyv\"askyl\"a, Finland. His research interests include modeling and performance enhancement of vehicular networks, physical layer security, ambient backscatter communications, and wireless power transfer. He was a recipient of the Outstanding Reviewer Award in 2017 from Elsevier.
\end{IEEEbiographynophoto}
%%%%%%%%%%%%%%%%%%%%%%%%

\begin{IEEEbiographynophoto}{Tapani Ristaniemi} received the M.Sc. degree in mathematics in 1995, the Ph.Lic. degree in applied mathematics in 1997, and the Ph.D. in wireless communications in 2000 from the University of Jyväskylä, Jyvaskyla, Finland. In 2001, he was appointed as a Professor with the Department of Mathematical Information Technology, University of Jyvaskyla. In 2004, he moved to the Department of Communications Engineering, Tampere University of Technology, Tampere, Finland,
where he was appointed as a Professor in wireless communications. Prof. Ristaniemi is currently a Consultant and a member of the Board of Directors of Magister Solutions Ltd. He is currently an Editorial Board Member of Wireless Networks and International Journal of Communication Systems.
\end{IEEEbiographynophoto}
%%%%%%%%%%%%%%%%%%%%%%%%

\end{document}